\documentclass{aastex631}
\usepackage[T1]{fontenc}
\usepackage{float}

\shorttitle{Long-period variability in SDSS Stripe 82 standards catalog}
\shortauthors{Fatović et al.}

\graphicspath{{./}{figures/}}

\begin{document}

\title{Detecting long-period variability in SDSS Stripe 82 standards catalog}

\author[0000-0003-1911-4326]{Marta Fatović}
\affiliation{Ru\dj er Bošković Institute, 
	Bijenička cesta 54,
	10000 Zagreb, Croatia}

\author[0000-0003-3710-0331]{Lovro Palaversa}
\affiliation{Ru\dj er Bošković Institute, 
	Bijenička cesta 54,
	10000 Zagreb, Croatia}

\author[0000-0001-6382-4937]{Krešimir Tisanić}
\affiliation{Ru\dj er Bošković Institute, 
	Bijenička cesta 54,
	10000 Zagreb, Croatia}

\author[0000-0003-1187-2544]{Karun Thanjavur}
\affiliation{Department of Physics \& Astronomy, University of Victoria,
	 3800 Finnerty Road,
	  Victoria, BC V8P 5C2, Canada}

\author[0000-0001-5250-2633]{Željko Ivezić}
\affiliation{Department of Astronomy and the DiRAC Institute, University of Washington,
	 3910 15th Avenue NE,
	  Seattle, WA 98195, USA}

\author[0000-0002-0786-7307]{Andjelka  B. Kova{\v c}evi{\'c}}
\affiliation{Department of Astronomy, Faculty of Mathematics, University of
	Belgrade, Studentski trg 16,11000 Belgrade, Serbia}
\affiliation{PIFI Research Fellow, Key Laboratory for Particle Astrophysics,
	Institute of High Energy Physics, Chinese Academy of Sciences,19B Yuquan Road,
	100049 Beijing, China}

\author[0000-0002-1134-4015]{Dragana Ili\'{c}}
\affiliation{Department of Astronomy, Faculty of Mathematics, University of
	Belgrade, Studentski trg 16,11000 Belgrade, Serbia}
\affiliation{Humboldt Research Fellow, Hamburger Sternwarte, Universit{\"a}t
	Hamburg, Gojenbergsweg 112, 21029 Hamburg, Germany}

\author[0000-0003-2398-7664]{Luka \v{C}.\ Popovi\'{c}}
\affiliation{Astronomical Observatory, Volgina 7, 11060 Belgrade, Serbia}
\affiliation{Department of Astronomy, Faculty of Mathematics, University of
	Belgrade, Studentski trg 16,11000 Belgrade, Serbia}
\affiliation{PIFI Research Fellow, Key Laboratory for Particle Astrophysics,
	Institute of High Energy Physics, Chinese Academy of Sciences,19B Yuquan Road,
	100049 Beijing, China}

\begin{abstract}
We report the results of a search for long-period ($100<P<600$ days) periodic variability in SDSS Stripe 82 standards catalog. The SDSS coverage of Stripe 82 enables such a search because
there are on average 20 observations per band in $ugriz$ bands for about 1 million sources, collected over about
6 years, with a faint limit of $r\sim22$ mag and precisely calibrated 1-2\% photometry.
We calculated the periods of candidate variable sources in this sample using the Lomb-Scargle periodogram and considered the three highest periodogram peaks in each of the $gri$ filters as relevant. Only those sources with $gri$ periods consistent within 0.1\% were later studied. We use the Kuiper statistic to ensure uniform distribution of data points in phased light curves. We present 5 sources with the spectra consistent with quasar spectra and plausible periodic variability. This SDSS-based search bodes well for future sensitive large-area surveys, such as the Rubin Observatory Legacy Survey of Space and Time, which, due to its larger sky coverage (about a factor of 60) and improved sensitivity ($\sim2$ mag), will be more powerful for finding such sources.

\end{abstract}

\keywords{Astronomy data analysis---Period determination---Time series analysis---Quasars }

\section{Introduction} \label{sec:intro}

Recent large-area time-domain sky surveys, such as the Optical Gravitational Lensing Experiment, \cite[OGLE,][]{1997AcA....47..319U,2003AcA....53..291U}, the Catalina Real-Time Transient Survey \citep[][]{2009ApJ...696..870D}, the Palomar Transient Factory \citep[PTF,][]{2009PASP..121.1395L}, Gaia \citep[]{2018A&A...616A...1G,2022arXiv220800211G}, and the Zwicky Transient Facility \citep[ZTF,][]{2019PASP..131a8002B} to name but a few, have shown the power of measuring variability of celestial sources for studying a variety of astrophysical phenomena. In this work we focus on one of the currently least well constrained poulations of astrophysical variability, the long-period, small-amplitude variability. The main reason why this domain of variability has not been studied in the past is the lack of adequate observational material: long observational baseline, photometric precision and depth over a large area of the sky are required to permit this kind of study.

There are several surveys that partially meet some the necessary requirements, however none of them bring them all together like Sloan Digital Sky Survey's (SDSS) time-domain survey in the Stripe 82 (S82) region (described in detail in Section~\ref{sec:data}). This survey has photometric precision and depth (faint limit $r\sim22$ mag) better than for example the Lincoln Near-Earth Asteroid Research \citep[LINEAR,][]{2000Icar..148...21S, 2013AJ....146..101P}, Palomar Transient Factory \cite[PTF,][]{2009PASP..121.1395L}, Gaia \citep{2018A&A...616A...1G} and Zwicky Transient Facility \citep[ZTF,][]{2019PASP..131a8002B, 2019PASP..131g8001G}. Also, with on average $\sim$20 epochs in the $gri$ bands, it has more epochs than, Pan-STARRS \citep[PS1,][]{2016arXiv161205560C, 2016ApJ...817...73H} and Dark Energy Survey \citep[DES,][]{2016MNRAS.460.1270D}.  The only surveys, other than SDSS S82, with which it is possible to study sources with a few percent change in brightness are Optical Gravitational Lensing Experiment \cite[OGLE,][]{1997AcA....47..319U,2003AcA....53..291U} and Kepler \citep{2011AJ....141..108C}. The disadventage of OGLE over SDSS is in the fact that OGLE only observes the inner galactic bulge and the Magellanic Clouds. Similarly, SDSS covers a longer time span and a larger area of the sky than Kepler. 

Our search for periodically variable objects in S82 was unbiased: we did not look for a  specific type of variability, but aimed to utilise S82’s long observational baseline to discover any type of long-period variability. This could include both galactic and extragalactic sources. Possible galactic long-period candidates were multiperiodic, semi-regular variable red stars such as the OGLE Small Amplitude Red Giants (OSARGs) \citep{2004MNRAS.349.1059W} and possibly some binary stars with edge-on orbits and nearly sinusoidal light curves. Periods of Mira stars also span the range of periods interesting to us, however, their large amplitudes would preclude them from becoming a part of the sample we analyzed, as large amplitude sources were already tagged in the S82 data by previous studies (see Section~\ref{sec:data}), and we specifically avoided those sources. It is possible, however, that some of the Miras in Stripe 82 were not detected in the earlier or our analysis. Both approaches depend on stable periodicity over the observational baseline for classification, which may not be the case for all Miras, who are known to exhibit period changes. We did not find any OSARGs and we could not confirm whether any OSARGs were indeed present in our sample as there is no overlap between OGLE surveys and the SDSS. We estimate that it would be difficult to identify OSARGs from the Stripe 82 data alone, given their multiperiodicity, small amplitudes and relatively sparse sampling of the Stripe 82 light curves. Extragalactic long-period variable candidates in this sample include quasars that may have a periodically varying optical light curve. The explanations of this behaviour include radio jet precession \citep[e.g.][]{2006IAUS..230..239R, 2011A&A...526A..51K}, tilted (warped) accretion disks \citep[e.g.][]{2014MNRAS.441.1408T}, tidal distruption events \citep[e.g.][]{2014ApJ...786..103L} and supermassive binary black hole systems \citep[e.g.][]{2015Natur.518...74G, 2008Natur.452..851V, 2018AAS...23110502L}.
Our final sample of 5 candidates with convincing periodic behaviour consists of sources that have spectra consistent with quasar spectra. 

Our paper is organized as follows: we describe the dataset used in our analysis in Section~\ref{sec:data}, report our analysis in Section~\ref{sec:analysis}, and discuss and summarize our results in Section~\ref{sec:results}. 
	
\section{SDSS Stripe 82 Imaging Data} \label{sec:data}

One of the largest regions on the sky with multi-band photometry precise to about 0.01 mag, faint
limit reaching $r\sim22$, and $>10$ observations per object is a 300 deg$^2$ region known as
the SDSS Stripe 82. Stripe 82 is a contiguous equatorial region that stretches between
$-60^{\circ}\leq{\rm R.A.}\leq60^{\circ}$ [20h to 4h], and $-1.266^{\circ}\leq{\rm Dec}\leq1.266^{\circ}$.
Following the initial concerted effort by the SDSS collaboration between 2001 and 2008 to map this
region repeatedly to a forecast imaging depth, $r \leq 22$, several other surveys in various wavebands
also have targeted this patch of sky to provide a rich multi-wavelength dataset suitable for a variety
of investigations. SDSS observations too have continued in this region \citep[e.g., the SDSS-II search for supernovae,][]{2008AJ....135..338F}, resulting in more epochs than initially planned.  

Data from the SDSS imaging camera \citep{1998AJ....116.3040G} are collected in drift-scan mode, sequentially in each of the five Sloan filters in the order $riuzg$.  The images that correspond to the same sky location in each of these five photometric bandpasses (these
five images are collected over $\sim$5 minutes, with an exposure time of 54 seconds for each band) are
grouped together for simultaneous processing as a {\it field}. A field is defined as a 36 seconds (1361 pixels)  
stretch of drift-scanning data from a single CCD column, referred as a {\it camcol}. Therefore, given this mode of data collection, the photometry in the $ugriz$ filters may be considered to be essentially simultaneous when interested in time scales of a day or longer.

\subsection{The 2007 SDSS Standard Star Catalog}

\defcitealias{Ivezi__2007}{I007}

The SDSS standard star catalog published by \cite{Ivezi__2007}, hereafter \citetalias{Ivezi__2007} , was constructed by averaging
multiple SDSS photometric observations (at least four per band, with a median of 10) in the $ugriz$ system.
The catalog includes 1.01 million presumably non-variable unresolved objects. The averaged measurements
for individual sources have random photometric errors below 0.01 mag for stars brighter than 19.5, 20.5, 20.5, 20,
and 18.5 in $ugriz$, respectively (about twice better than for individual SDSS runs).  

The 1.01 million standard stars in the \citetalias{Ivezi__2007} catalog were selected as non-variable sources by requiring that for each source their $\chi_{dof}^2<3$ ($\chi^2$ per degree of freedom) in each of the $gri$ bands, under the assumption of constant brightness.
In addition, about 67,000 rejected light curves showed clear variability, with $\chi_{dof}^2$ per degree of freedom exceeding 3,
and the root-mean-square (rms) variability exceeding 0.05 mag, in both $g$ and $r$ bands. The behavior of
such obviously variable sources, dominated by RR Lyrae stars and quasars, was analyzed in detail by
\cite{2007AJ....134.2236S} and was not a part of this work.

Given that here we are interested in finding long periodic variability considering light curves for sources listed in the standard star catalog (i.e., those with $\chi^2<3$), the assumption is that the amplitudes of the sources found will be small. In other words, we aim to use periodogram analysis to uncover long-period variable sources with possible small-amplitudes that were {\it not} recognized as variable by \citetalias{Ivezi__2007}. Furthermore, our final dataset includes more data than originally used by \citetalias{Ivezi__2007}, as described next.

\subsection{Post-2007 SDSS data \label{ssec:DR15}}

Recently, \cite{2021MNRAS.505.5941T} extended light curves assembled by \citetalias{Ivezi__2007} with SDSS data obtained after
2007. Using the SDSS Data Release 15 (DR15) as available in April 2019 \citep{Blan17}, they constructed
light curves for stars from the standard star catalog with about twice as many data points as available to
\citetalias{Ivezi__2007} (about 20 on average and extending to 50 depending on the position within Stripe 82, see their Fig. 1).
We note that DR15 does not include runs from the SDSS-II supernovae surveys \citep{2008AJ....135..338F},
which are typically of lower photometric quality. For more details, such as photometric recalibration of this
new dataset, we refer the reader to \cite{2021MNRAS.505.5941T}.

The final dataset, consisting of all the light curves in the five $ugriz$ filters for the presumably 1,001,592
non-variable standard stars from the \citetalias{Ivezi__2007} catalog resulted in $\sim$20 GB of tabular data. This catalog is the starting point of our search for periodic variability. To make file search and access fast, the data have been organized into sub-directories, each spanning 1 $\deg$ in RA, and 0.1 $\deg$ in Dec  (a "poor-man's" two-dimensional tree structure). The light curves from this catalog are \href{http://faculty.washington.edu/ivezic/sdss/catalogs/stripe82.html}{publicly available}\footnote{http://faculty.washington.edu/ivezic/sdss/catalogs/stripe82.html}.
\section{Light Curve Analysis \label{sec:analysis}}

Since surveys mentioned earlier already explored the fast and large-amplitude variability in the S82 region, we focus on the long-period, small amplitude optical variations. We also show that the theoretical limit to which variability can be plausibly discovered with our data is at the level of A$\sim$0.03 mag in SDSS r-band. 

We start with a sample of 1,001,592 light curves and select a subset of	143,505 light curves that have at least N = 25 observational epochs after filtering out epochs with likely spurious photometry. We identify the latter by having an epoch \textit{gri} magnitude value outside the interval between 11 and 23 mag, or unrealistically small photometric errors	(smaller than 0.0001 mag), or photometric errors larger than 0.2 mag. This minimum number of data points, together with photometric errors, implies the
minimum variability amplitude that we can detect.
In case of a source with constant brightness, and a light curve with at least 10 data
points and reliably estimated photometric errors, the $\chi_{dof}^2$ distribution can be
approximated as a Gaussian (normal) distribution with an expectation value of 1 and a
standard deviation of $\sqrt{2/N}$. For illustration, for $N=25$ only about one non-variable source per
thousand would have $\chi_{dof}^2 > 1.85$. Let us assume a hypothetical population of
variable sources that represent, say, 0.01\% of the sample, and have a variability amplitude
$A$ that results in $\chi_{dof}^2$ distribution (under the assumption of no variability) with
a median of 1.85. If now a subsample of all sources is selected with $\chi_{dof}^2 > 1.85$,
it will include 95\% of false positives (randomly scattered non-variable sources) and 5\% of
truly variable sources (i.e. 50\% of all truly variable sources). Although the contamination
rate is 19:1, let us assume here that a follow-up analysis can further ``clean the sample''.

What is the variability amplitude $A$ that results in $\chi_{dof}^2$ distribution with a
median of 1.85? It is easy to show that for a well sampled light curve following
$y(t) = A \rm{sin}(\omega t)$ with homoscedastic Gaussian errors with standard deviation
$\sigma$, the variance is $V =\sigma^2+A^2/2$ \citep{2019sdmm.book.....I}.
The requirement that $\chi_{dof}^2 = 1.85$ yields a minimum detectable amplitude, $A = 2.9\sigma/N^{1/4} = 1.3 \sigma$ for our requirement of N $\geq$ 25.  
Therefore, we expect that the completeness for truly variable sources when using selection
$\chi_{dof}^2 > 1.85$ is at least 50\% for $A \sim 0.03$ (using $\sigma \sim0.02$ mag)
and higher for larger values of $A$. In conclusion, we expect that our dataset is sensitive to
amplitudes $A>$ 0.03 mag. According to Kepler results discussed by \cite{2014ApJ...796...53R},
about 3\% of stars are expected to show variability with $A>0.02$ mag (including non-periodic
variables and for all values of period for periodic variables).

Search for periodicity should produce $\chi_{dof}^2 \sim 1$ in case of truly periodic variables and
thus yield a reduction of $\chi_{dof}^2$ by about a factor of two (or more in case of amplitudes
larger than $\sim$0.03 mag). This reduction of $\chi_{dof}^2$ is the main method we employ
here, but we supplement it with additional tests because the theoretical analysis above is
sensitive to period aliasing and non-Gaussian behavior of photometric errors. 

In the following subsections we outline our selection procedure.

\subsection{Lomb-Scargle Periodogram analysis \label{sec:initial} }

For each of the 143,505 light curves, we computed Fast Lomb-Scargle periodograms \citep{1989ApJ...338..277P} using {\tt astropy} implementation of the algorithm \citep{astropy-1, astropy-2}. The spacing of the search grid was selected automatically by the {\tt autopower} method, which takes the user-supplied minimum and maximum frequencies as the input parameters ($f_{min}=1/600$ days$^{-1}$, $f_{max}=1/2$ days$^{-1}$) and adjusts the grid spacing according to the number of observations of the given source and the length of the observational baseline. This method will assign five grid points across each significant periodogram peak. 
The upper period search boundary was set to 600 days in order to limit the computational effort required to search for periods of approximately more than a hundred thousand sources. Lower period boundary was set to 2 days since periods of $P<1$ day would fall within the (short period) variable stars which have already been studied in detail in the SDSS S82 region by previous studies. Also, $P\sim1$ day is strongly aliased so we moved lower limit up by a factor of two. 
Since $gri$ bands typically have the highest photometric signal-to-noise ratio (i.e. smallest photometric errors), we considered periodograms calculated using only the $gri$ data. We added a systematic photometric error of 0.01 mag in quadrature to photometric errors (uncertainties) reported by the SDSS photometric pipeline to avoid (rare) cases of unrealistically small reported errors.

In each band, we retained periods corresponding to the three highest periodogram peaks in order to have more options when validating possible contamination by aliasing. Taking more than three periodogram peaks would have typically sampled noise instead of the real signal. Given these nine periods per source, we select light curves with plausible periodic variability by requiring at least one period to be in common to each of the \textit{gri} bands (to within 0.1\%). Applying these constraints resulted in 1,078 unique sources with 2,135 corresponding periods.

\subsection{$\chi^{2}$ analysis \label{sec:chi2}}

For the selected 1,078 candidate periodic variable sources, we constructed phased light curves folded with the corresponding 2,135 periods. Then we checked how well the data agree with the sinusoidal light curve model by calculating $\chi^{2}$ and Lomb-Scargle periodogram in the following way:
\begin{equation}
P_{LS}=1-\frac{\chi^{2}_{per}}{\chi^{2}_{const}}.
\end{equation}
Values $\chi^{2}_{per}$ and $\chi^{2}_{const}$ were calculated as:
\begin{equation}
\chi^{2}_{per}=\sigma_{G}\left(x_{per}\right)^{2}, \mbox{ where: } x_{per} = \frac{mag-model}{magErr},
\end{equation}
\begin{equation}
\chi^{2}_{const}=\sigma_{G}\left(x_{const}\right)^2, \mbox{ where: } x_{const}=\frac{mag- \langle mag \rangle}{magErr}.
\end{equation}
The used function $\sigma_{G}(x)=0.7413 \cdot (x_{75} - x_{25})$ is the normalized interquartile range of the observed source magnitudes. We demanded $P_{LS} > 0.5$, in other words we are requiring the improvement of the $\chi^{2}$ by the factor of 2. This requirement reduces the candidate sample to 342 unique sources with 601 periods.

Given the simple sinusoidal model that was used, it is possible that some sources with light curves significantly more complex than this model were rejected. However, during the analysis we also performed a visual verification of the candidate light curves and found no evidence of sources with more complex light curves.

\subsection{Period analysis \label{sec:per_an}}
\begin{figure}

	\centering
	\resizebox{0.6\hsize}{!}{\includegraphics{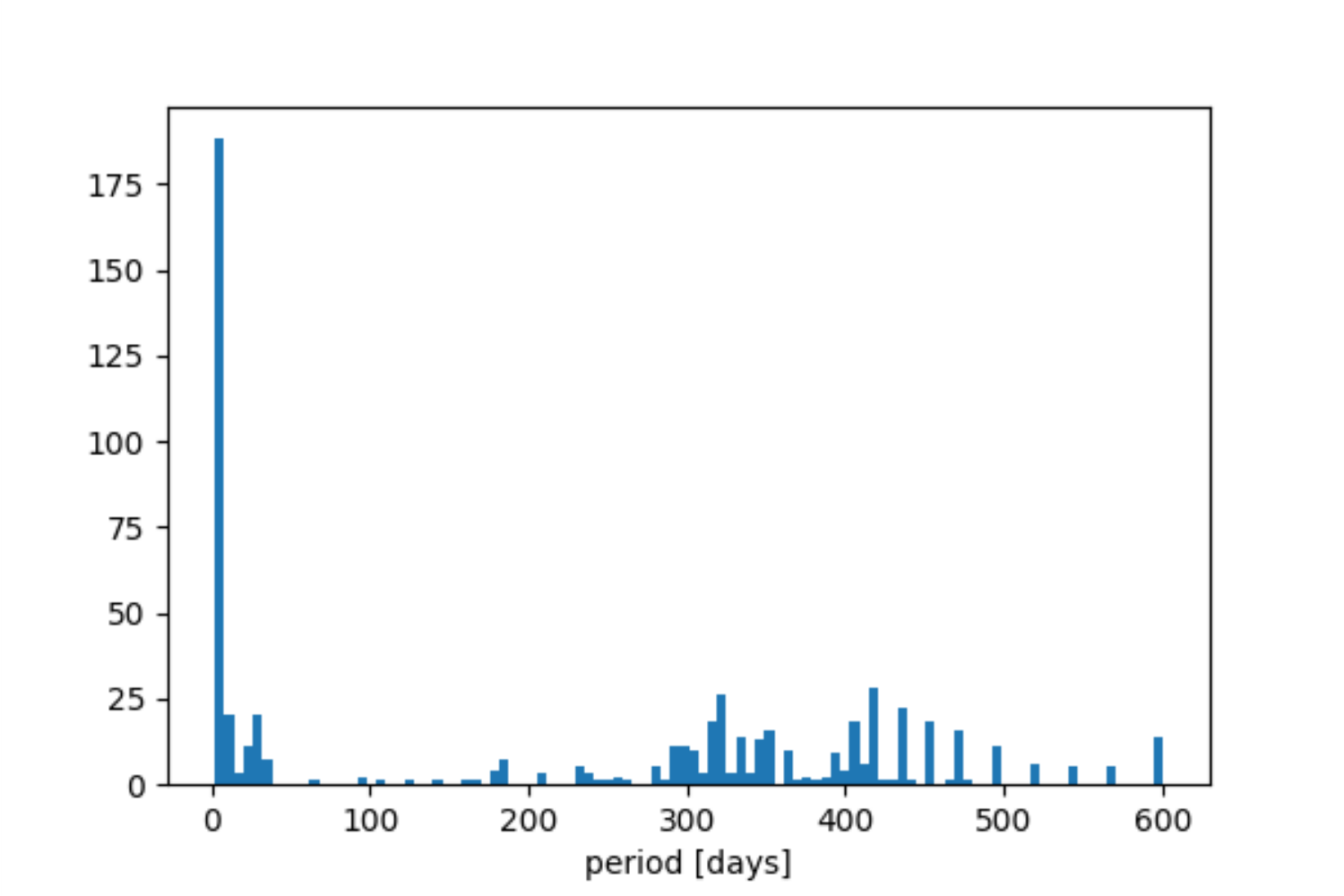}}

	\caption{Histogram of periods for 342 sources with 601 corresponding periods.}
	\label{fig:per_hist}
\end{figure}

From the period histogram in Figure~\ref{fig:per_hist} it is clear that there are a few period values that are recurring in Lomb-Scargle fits for different sources. It is natural to suspect that these periods are in fact aliases. Alternatively, this could mean we might have missed some true periods due to the coarseness of the grid selected by the \textit{autopower} method applied by the Lomb-Scargle routine.

\begin{figure}
	{ \hspace{+1cm} \centering   \resizebox{ 0.67 \hsize}{!}{\includegraphics{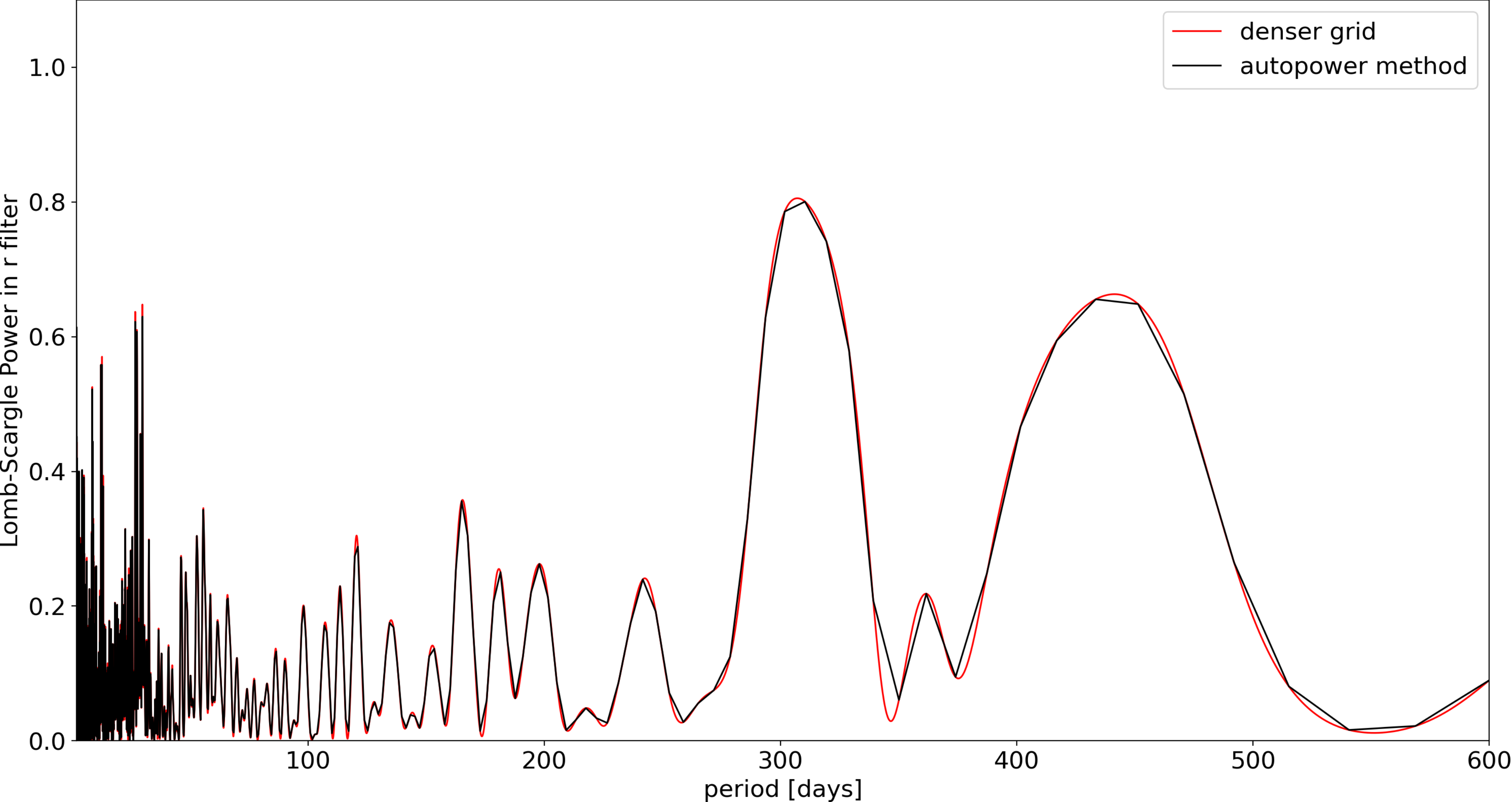}}
	\caption{A typical example of a Lomb-Scargle periodogram for a candidate variable. Figure shows periodogram obtained using \textit{autopower} method and the one calculated with a denser grid. \label{fig:LS-both}}}
	
\end{figure}

To investigate the impact of using a sparse grid on the calculation of the periods, we re-ran the Lomb-Scargle period search on a denser grid.  Since our sample was now much smaller than the one we started with, we were able to do it without demanding higher computational power. We used the same frequency range as before ($f_{min}=1/600$ days$^{-1}$, $f_{max}=1/2$ days$^{-1}$) and the grid was defined in the following way: for periods between 2 and 8 days, the time spacing ($\Delta t$) was 1 minute, for periods between 8 and 32 days $\Delta t$ was 3 minutes, for periods between 32 and 128 days $\Delta t$ was 10 minutes and for periods between 128 and 600 days $\Delta t$ was 1 hour. By comparing the results of the both period estimation runs, we were able to ascertain that no significant periodogram peaks were missed in the run with sparser period search grid. The Figure \ref{fig:LS-both} shows an example of the comparison of periodograms obtained with a denser grid and with the \textit{autopower} method. It is clear that there is no significant difference between them.

We discarded all of the (clustered) periods that can fit within a 0.1 day bin. Considering that these periods agree within 0.1 days, it is likely that they are aliases.
That left us with 69 sources with 82 periods.
For further analysis, we retained only objects with periods longer than 100 days. This boundary was applied in order to consider only long-period variability. Also, Figure~\ref{fig:per_hist} shows that most periods up to 100 days are grouped around small values (around zero). This is probably due to the 1-day aliasing caused by the Earth's rotation. Furthermore, we have already mentioned that the catalog was previously cleaned of all short-period variable sources, so such cases were not considered in this work. 

This constraint left us with 45 unique objects and 58 corresponding periods. 

\subsection{Phase distribution analysis \label{sec:prelim}}

Further analysis was based on a requirement that every candidate should have a fairly uniform distribution of observations with respect to the phase. We base this requirement on the assumption that it is unlikely that for a significant fraction of the observed sources the cadence of the survey will be matched to the period of a particular source in a way that would produce observations that are always near the same point (or a few points) in the phased light curve. Indeed, given the uniformity of the cadence across all observed S82 fields, having a significant fraction of objects with clumped observations in the phased light curve would imply that those objects would have a common frequency $f_c \in f/n$, where $f$ is the period of the object and $n$ is a small integer. We note that with this requirement we may lose a fraction of good candidates, but since we are more concerned about purity of the sample than its completeness, every source whose phase-folded light curve exhibit clumping of observations was characterized as a false positive and eliminated from further analyses.

We quantify the uniformity of the distribution of the observed phases for each source using the Kuiper’s
statistic. This statistic compares an empirical cumulative distribution function defined on a circle with the expected cumulative distribution function for a uniform distribution (CDF(x) = x). The maximum deviations are given as:
\begin{equation}
D = max(CDF_{data} - CDF_{unif}) + max(CDF_{unif} - CDF_{data}) .
\end{equation}
We calculated the probability of obtaining the calculated D by a random fluctuation for a uniform distribution. When the probability is very small we conclude that we have non-uniform phase coverage, and therefore reject the corresponding period. D \% for a uniform distribution, is a function of sample size, N: $ D \% = \frac{C}{N^{p}} $, where for the case of 99.9 \% we used $C = 2.04$ and $p = 0.486$, obtained by fitting this functional form to numerical simulations of draws from uniform distributions of varying size $8 < N < 10^{4}$, and valid for $0.95 < D\% < 0.999$.
We only retained those that have $D<D_{99.9 \%}$ in each of the $gri$ bands. After this filter, our sample decreased to 28 sources with 33 periods.
\\

\subsection{Source type determination and the determination of period uncertainties \label{sec:finfilt}}

\begin{figure}
	{ \centering   \resizebox{ 0.7 \hsize}{!}{\includegraphics{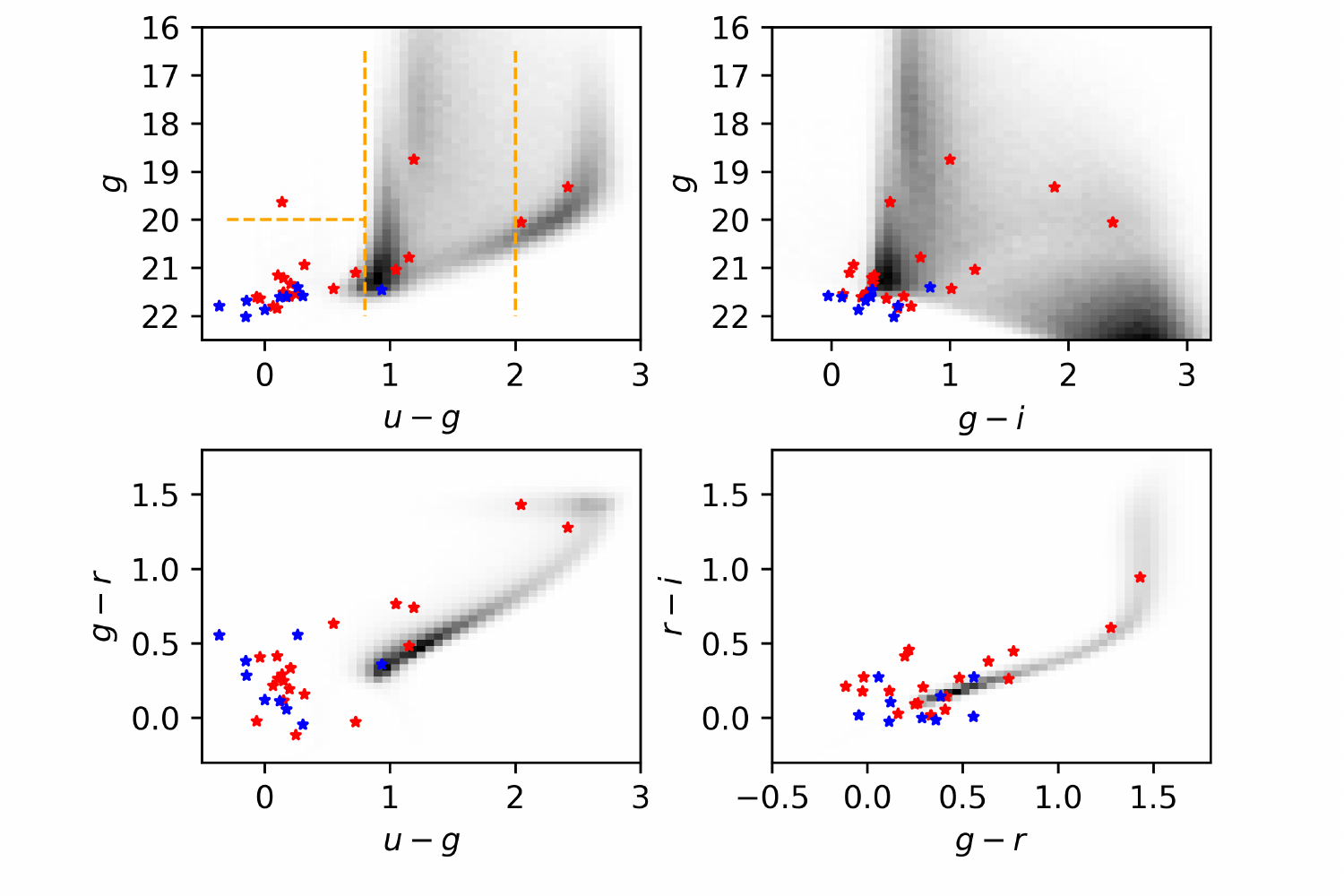}}
	\centering

	\caption{Color-magnitude and color-color diagrams for 28 candidates. Sources with period $P<365$ days are plotted as red and those with $P>365$ days are plotted as blue symbols. The gray background represents the distribution of all objects from the catalog. \label{fig:CM_CC} }}
\end{figure}

In order to better characterise our final list of periodically variable sources, at this stage we used color-color and color-magnitude diagrams shown in Figure~\ref{fig:CM_CC}. The clustering observed in the upper left panel was a motivation for introducing 4 subgroups described in Table~\ref{tab:regions}. For each source we introduce an ID based on the location in the $g$ vs $u-g$ plot in Figure \ref{fig:CM_CC}.
\begin{table}[htbp]
	\caption{The division of the sample in 4 subgroups. \label{tab:regions} }        
	\centering           

	\begin{tabular}{l l l}     
		\hline\hline                      
		Regions & Types & ID \\   
		\hline                                  
		$i)$ u-g > 2.0 & “late-type” K and M stars  & 0-1  \\     
		$ii)$ 0.8 > u-g < 2.0 & “intermediate-type” F and G stars  & 2-5      \\
		$iii)$ u-g < 0.8 and g < 20 & quasars; possible presence of stars & 6        \\
		$iv)$ u-g < 0.8 and g > 20& quasars & 7-27        \\
		\hline                                            
	\end{tabular}
	\tablecomments{Using the logic of this partition, we introduced new names for each source which are given in column ID.}
\end{table}

\subsubsection{Monte - Carlo simulations \label{MC}}
We also preformed Monte - Carlo simulations to calculate the uncertainty of the estimated periods. The procedure and results are given in Appendix~\ref{Sec:AppB}. Using the results obtained with this method, we further filtered our sample. Requiring the agreement of previously calculated periods and those calculated by this method within the obtained uncertanty ($|P-P_{gatspy}| < \sigma_{P}$), our sample is reduced to 9 sources with 10 corresponding periods. 

\subsubsection{One-year period alias filtering \label{sec:alias}}

As a final filter, we check if the calculated periods are a real signals, or just aliases of one year. We used:

\begin{equation}  \label{eq:alias}
P_{a} = 365 \pm \frac{k}{n} \cdot 365; \ \ k = 1,2; \ n = 1,2,3,4,5,
\end{equation}
to identify possible one-year aliases. If the calculated period P is within the range $P \pm \sigma_{P}$, where $\sigma_{P}$ is the period uncertainty obtained with simulations (the procedure is explained in Appendix~\ref{Sec:AppB}), we consider it an alias. Application of this filter left us with the final sample of 5 candidate periodically variable sources with 6 corresponding periods.

\subsubsection{SDSS Cutouts \label{sec:cutouts}}
Since apparent variability in magnitude can be caused by a potential nearby source, we checked SDSS cutouts for each of the final 5 sources. By inspecting the images we confirmed that no candidates are affected by blending. An acceptable variable candidate should be isolated from all of the objects in its neighborhood. 

\subsubsection{Light Curves \label{sec:LC}}
In order to investigate the behavior of the light curves of each source at times different from those of the SDSS observations, we also searched the databases of other surveys. We found information about the light curves for the final 5 sources in ZTF DR11 \citep[][]{2019PASP..131a8003M} and Pan-STARRS \citep[PS1, ][]{2016arXiv161205560C,2020ApJS..251....7F}. We simultaneously plotted SDSS, ZTF and PS1 light curves in corresponding filters, along with the model.
Our minimum requirement was that the apparent variability is confirmed in later times at least in one of the filters for at least one of the additional surveys. 
Apparent variability in additional surveys was confirmed for all 5 sources, which reduces the likelihood that the data from the original survey (SDSS) is bad data and that the detected (periodic) variability in SDSS is due to the random fluctuation.
 
\section{Results} \label{sec:results}
\subsection{A summary of the applied filters and their effects on the selection. \label{sec:FiltSummary}}

Here we summarize the effect of each of the filters above on our selection procedure. We started with 143,505 light curves that have at least N =25.                  
\begin{itemize}

	\item  each of the $gri$ periods agree to within 0.1\% $\rightarrow$ 1,078 sources with 2,135 corresponding periods
	\item  $\chi^{2}$ and $P_{LS} > 0.5$ analysis $\rightarrow$ 342 sources with 601 corresponding periods
	\item  "repeated" periods ($\Delta P = 0.1$ days) and periods  100 days < P < 600 $\rightarrow$ 45 sources with 58 corresponding periods
	\item  Kuiper statistic: $D<D_{99.9\%}$ $\rightarrow$ 28 sources with 33 corresponding periods
	\item  comparison with Monte-Carlo results $\rightarrow$  9 sources with 10 corresponding periods
	\item  one-year period alias $\rightarrow$ 5 sources with 6 corresponding periods
	\\
	\\
	Additional checks: 
	\item  blending $\rightarrow$ no sources were discarded
	\item  comparison to more recent light curves $\rightarrow$ no sources were discarded
\end{itemize}
In the next section we proceed to analyze the remaining candidates. 

\subsection{Plausible periodically variable candidates}

\begin{figure}
	{	\resizebox{0.9\hsize}{!}{\includegraphics{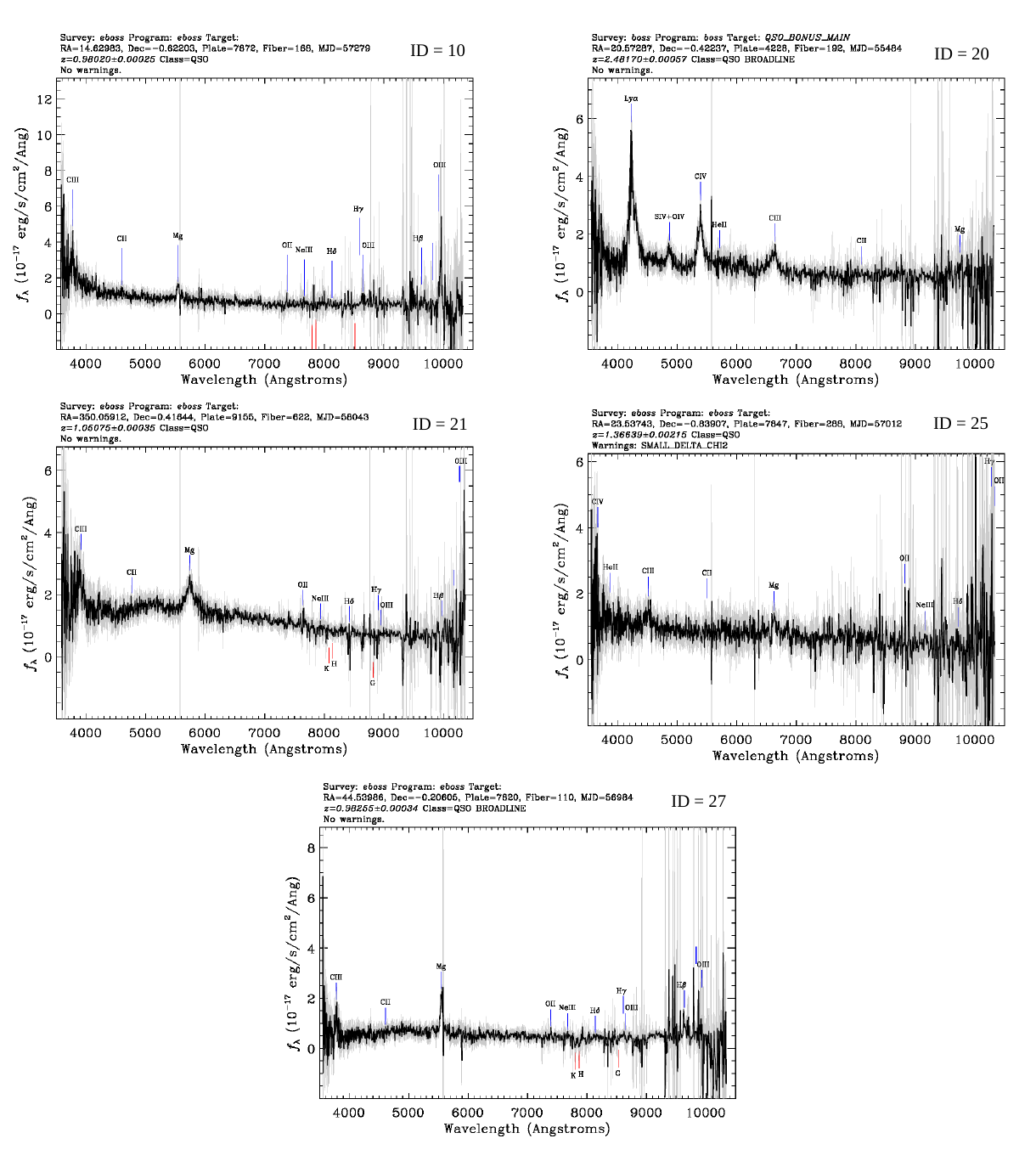}}
		\caption{SDSS spectra for 5 periodically variable sources. All of them are consistent with quasar spectra.}
		\label{fig:spec}}
\end{figure}

In order to better characterize our final sample we searched the SDSS DR16 online database for spectra of the candidate sources. We find that all 5 of the final candidates have a SDSS \citep[][]{2013AJ....145...10D, 2016AJ....151...44D} spectra consistent with quasar spectra. Their SDSS spectra are shown in Figure~\ref{fig:spec}.

All 5 sources have phased light curves with a uniform distribution of data points along the light curve, are well isolated from other sources and their (periodic) variability at later times is supported by either ZTF or PS1 survey data. The figures summarizing light curves of each of the sources, along with SDSS photometry cutouts are given in Appendix~\ref{sec:AppB}. 

Large uncertainties of the ZTF photometry prevent us from making any definitive arguments in case of the sources with IDs 10 and 25. However, PS1 photometry gives us sufficient variability confirmation in at least one of the bands. Source ID = 10 (Figure~\ref{fig:qsos_1} a)) has the smallest amplitude ($A \sim 0.2$ mag) of all of the final candidates. ID = 25 (Figure~\ref{fig:qsos_2} d) and e)) source has two reported periods that passed our selection criteria (P$_{1}$ = 300 d, P$_{2}$ = 299 d). Since they are within the uncertainty given in Table~\ref{Tab:tab2} ($|P_{1}-P_{2}| < \sigma_{P}$), we conclude that the same period is recovered by both methods (essentially 300 days).

Sources ID = 20, 27 boast a robust periodic variability confirmation by ZTF and PS1. Also, in Figure~\ref{fig:qsos_1} b) and Figure~\ref{fig:qsos_2} f) can be seen how ZTF amplitudes do not exclude SDSS-based model as being incorrect.

We single out a quasar with ID = 21 (Figure~\ref{fig:qsos_1} c)) and P = 278 days as the most interesting one in this sample. Besides the optical variability detected in this work, this source also has Chandra X-ray catalog variability flag set to 1 (source displays flux variability within one or more observations, or between observations, in one or more energy bands) \citep{2010ApJS..189...37E}. For this source we performed an additional analysis of the periodicity with the 2D hybrid model \citep[][]{2018MNRAS.475.2051K, 2020OAst...29...51K}, and found that this period and period calculated with the method explained in Section~\ref{sec:initial} are in excellent agreement. A detailed description of the procedure is given in Appendix~\ref{Sec:AppC}. 
\\
\section{Summary and Conclusions} \label{sec:concl}
We report the results of a search for long-period variability in the SDSS Stripe 82 region. Starting with a sample of about million presumably constant sources we selected 5 as plausible candidates with periodic variability after a thorough analysis of their light curve characteristics and ancillary data. 
Our final sample consists of 5 sources with SDSS spectra consistent with quasar spectra. We consider our final candidates to be very likely to have periodic nature, although we cannot completely rule out the possibility of stochastic behavior causing the (apparently) periodic signal. 

Determining the cause(s) of the variability of our candidates is not possible without additional data. Therefore, we singled out one quasar (ID=21) with a period of $P=278$ days whose periodic variability could not be excluded by ZTF and PS1, while its X-ray variability is detected by Chandra. Additional tests described in Appendix~\ref{Sec:AppB} and \ref{Sec:AppC} included period estimations by a different models and used simulations to confirm reliability of the obtained results. Our simulations could not reject the assumption of periodicity of this quasar since its P-value is $\sim 60\%$. In order to further investigate the nature of its variability and possibly improve our approach to photometric detection of such sources in the context of detection exotic objects, and in particular gravitational wave emitters, we will be shortly obtaining Gemini GMOS-N spectra of the source.

We hope that this study provides to be useful in uncovering some of the long-period phenomena that may have been missed by the extant optical surveys. Although we initially expected long-period, low-amplitude stars in our final sample, we did not find them. Instead we found 5 excellent candidates for periodic variability displayed by spectroscopically confirmed quasars.

\newpage
\begin{acknowledgments}
We are grateful to John Tonry for providing to us ATLAS photometry for several stars. \v{Z}I acknowledges
hospitality by the Ru\d er Bo\v{s}kovi\'{c} Institute. 
\\
We acknowledge the support of the Center of Advanced Computing and Modelling, University of Rijeka (HPC Bura) for providing computing resources.
\\
This work is financed within the Tenure Track Pilot Programme of the Croatian Science Foundation and the Ecole Polytechnique Fédérale de Lausanne and the Project TTP-2018-07-1171 Mining the variable sky, with the funds of the Croatian-Swiss Research Programme.
 \\
Funding for the SDSS and SDSS-II has been provided by the Alfred P. Sloan Foundation, the Participating
Institutions, the National Science Foundation, the US Department of Energy, the National Aeronautics and 
Space Administration, the Japanese Monbukagakusho, the Max Planck Society, and the Higher Education Funding Council for England. The SDSS Web site is http://www.sdss.org.
SDSS-IV is managed by the 
Astrophysical Research Consortium 
for the Participating Institutions 
of the SDSS Collaboration including 
the Brazilian Participation Group, 
the Carnegie Institution for Science, 
Carnegie Mellon University, Center for 
Astrophysics | Harvard \& 
Smithsonian, the Chilean Participation 
Group, the French Participation Group, 
Instituto de Astrof\'isica de 
Canarias, The Johns Hopkins 
University, Kavli Institute for the 
Physics and Mathematics of the 
Universe (IPMU) / University of 
Tokyo, the Korean Participation Group, 
Lawrence Berkeley National Laboratory, 
Leibniz Institut f\"ur Astrophysik 
Potsdam (AIP),  Max-Planck-Institut 
f\"ur Astronomie (MPIA Heidelberg), 
Max-Planck-Institut f\"ur 
Astrophysik (MPA Garching), 
Max-Planck-Institut f\"ur 
Extraterrestrische Physik (MPE), 
National Astronomical Observatories of 
China, New Mexico State University, 
New York University, University of 
Notre Dame, Observat\'ario 
Nacional / MCTI, The Ohio State 
University, Pennsylvania State 
University, Shanghai 
Astronomical Observatory, United 
Kingdom Participation Group, 
Universidad Nacional Aut\'onoma 
de M\'exico, University of Arizona, 
University of Colorado Boulder, 
University of Oxford, University of 
Portsmouth, University of Utah, 
University of Virginia, University 
of Washington, University of 
Wisconsin, Vanderbilt University, 
and Yale University.
\\
Based on observations obtained with the Samuel Oschin Telescope 48-inch and the 60-inch Telescope at the Palomar Observatory as part of the Zwicky Transient Facility project. ZTF is supported by the National Science Foundation under Grant No. AST-2034437 and a collaboration including Caltech, IPAC, the Weizmann Institute for Science, the Oskar Klein Center at Stockholm University, the University of Maryland, Deutsches Elektronen-Synchrotron and Humboldt University, the TANGO Consortium of Taiwan, the University of Wisconsin at Milwaukee, Trinity College Dublin, Lawrence Livermore National Laboratories, and IN2P3, France. Operations are conducted by COO, IPAC, and UW.
\\
The Pan-STARRS1 Surveys (PS1) and the PS1 public science archive have been made possible through contributions by the Institute for Astronomy, the University of Hawaii, the Pan-STARRS Project Office, the Max-Planck Society and its participating institutes, the Max Planck Institute for Astronomy, Heidelberg and the Max Planck Institute for Extraterrestrial Physics, Garching, The Johns Hopkins University, Durham University, the University of Edinburgh, the Queen's University Belfast, the Harvard-Smithsonian Center for Astrophysics, the Las Cumbres Observatory Global Telescope Network Incorporated, the National Central University of Taiwan, the Space Telescope Science Institute, the National Aeronautics and Space Administration under Grant No. NNX08AR22G issued through the Planetary Science Division of the NASA Science Mission Directorate, the National Science Foundation Grant No. AST-1238877, the University of Maryland, Eotvos Lorand University (ELTE), the Los Alamos National Laboratory, and the Gordon and Betty Moore Foundation.
\\
This work has made use of data from the European Space Agency (ESA) mission
{\it Gaia} (\url{https://www.cosmos.esa.int/gaia}), processed by the {\it Gaia}
Data Processing and Analysis Consortium (DPAC,
\url{https://www.cosmos.esa.int/web/gaia/dpac/consortium}). Funding for the DPAC
has been provided by national institutions, in particular the institutions
participating in the {\it Gaia} Multilateral Agreement.
\\
This research has made use of the SIMBAD database,
operated at CDS, Strasbourg, France 
\\
Some of the data presented in this paper were obtained from the Mikulski Archive for Space Telescopes (MAST) at the Space Telescope Science Institute. The specific observations analyzed can be accessed via \dataset[DOI]{https://doi.org/10.17909/s0zg-jx37}.
\\
 \textit{Zwicky Transient Facility} data were extracted through the \cite{https://doi.org/10.26131/irsa538} portal.
\\
D.I., A.B.K, and L.\v C.P. acknowledge funding provided by the University of
Belgrade - Faculty of Mathematics (the contract 451-03-68/2022-14/200104),
Astronomical Observatory Belgrade (the contract 451-03-68/2022-14/ 200002),
through the grants by the Ministry of Education, Science, and Technological
Development of the Republic of Serbia. D.I. acknowledges the support of the
Alexander von Humboldt Foundation. A.B.K. and L.{\v C}.P thank the support by 
Chinese Academy of Sciences President's International Fellowship Initiative
(PIFI) for visiting scientist.
\end{acknowledgments}
\newpage
\begin{acknowledgments}
\facilities{SDSS, ZTF, Pan-STARRS, Chandra}
\software{numpy \citep{numpy}, matplotlib \citep{matplotlib}, scipy \citep{scipy}, 
	astropy \citep{astropy-1, astropy-2}, astroML \citep{2012cidu.conf...47V}}
\end{acknowledgments}

\newpage
\appendix

\section{Final candidates} \label{sec:AppA}
In Table \ref{Tab:tab1}, we provide the spectral type as determined by the SDSS DR16 pipelines and a summary of photometric characteristics derived from the S82 photometry for the the
final five candidates.
\noindent

\subsection{Table of final candidates}

\begin{table}[htbp]     
 
	\centering  
	
	\caption{Table of 5 long-period small-amplitude variability candidates. \label{Tab:tab1}  }  
	{ \hspace{-3cm} \centering     \resizebox{1.1\columnwidth}{!}  {
			\begin{tabular}{cccccccccccccccc}   
				\hline\hline      
				objID&ID&SPECTRAL \ TYPE&ra&dec&period&u&g&r&i&z&u-g&g-r&Amp(g)&Amp(r)&Amp(i) \\
				\hline
				1237666338651242745&10&QSO&14.629833&-0.622042&297.1988&21.30&21.18&20.93&20.84&20.4&0.12&0.25&0.28&0.27&0.22\\
				1237666338653864147&20&QSO&20.57287&-0.422368&466.3183&21.69&21.39&21.26&21.36&20.68&0.30&0.13&1.22&1.08&0.97\\
				1237666408438038843&21&QSO&350.059098&0.416449&277.8278&21.45&21.51&21.13&21.08&20.76&-0.06&0.38&0.63&0.52&0.44\\
				1237663783133970736&25&QSO&23.53744&-0.839078&299.7779&21.90&21.79&21.38&21.21&20.58&0.11&0.41&0.34&0.33&0.29\\
				1237666300553068800&27&QSO&44.539858&-0.206044&479.5917&21.81&21.73&21.43&21.32&20.71&0.08&0.30&1.13&0.96&0.82\\
				\hline                 
	\end{tabular}     }}

	\tablecomments{\textit{Columns :} objID: SDSS DR9 object ID; ID: ID given on the basis of the position of an object in color-magnitude diagram described in Table~\ref{tab:regions}; TYPE: Classification from SDSS DR16; ra / dec: coordinates; period: period of an object (in days) calculated as described in Section~\ref{sec:analysis} ; $u$/$g$/$r$/$i$/$z$: weighted mean of magnitude in each filter; $u$-$g$/$g$-$r$: mean colors; Amp($g$/$r$/$i$): the difference between the maximum and minimum of the sinusoidal model.}

\end{table}

\subsection{Representative figures of final candidates} \label{sec:AppB}

Here we summarize all the plausible periodically variable candidates represented by two panels that were used during our analyses.

Each figure contains:
\begin{itemize}
	\item Left column: SDSS cutouts described in Section~\ref{sec:cutouts},
	\item Middle column: light curves in $ugriz$ bands with light-
	curve model derived using SDSS data. In addition to
	SDSS data, ZTF and PS1 data are also plotted in the
	figure to check whether the variability can be confirmed
	at later times with additional surveys. To account for the
	difference between the instruments, a simple magnitude
	correction was applied to each of the $gri$ filters:
	
	\begin{equation}
	SURVEY_{corr} = SURVEY + \langle SURVEY \rangle  - \langle SDSS \rangle,
	\end{equation}
	
	\noindent where $SURVEY_{corr}$ is the corrected magnitude, i.e. magnitude shifted by the median difference between the survey in question and the SDSS ($\langle SURVEY \rangle - \langle SDSS \rangle$).	
	
	In some cases few ZTF and PS1 data points are not visible in the figures because their difference with respect to the model light curve is too large. In order to preserve
	convenient scaling of the plots, we left these outlying
	values outside the range of the plots and instead
	designated them with an arrow on the left-hand side in
	some of the panels in the middle column. The arrows are color-coded according to the survey and the arrows' length ($l$) is proportional to the number of data points ($N$) outside the ordinate range, with: $l = N \cdot 0.05 + 0.03$. Furthermore, given the shallower ZTF limiting magnitudes with respect to Pan-STARRS and SDSS ($~$20.5 mag), and consequently larger scatter at the ZTF's faint end, we binned the ZTF data in 7-day intervals and calculated a weighted mean magnitude for each bin in order to improve photometric precision and enhance the legibility of the plots.  
	\item Right column: phased light curves in $ugriz$ bands.
\end{itemize}

\newpage

\begin{figure*}[h!]
{	\centering
	\gridline{\fig{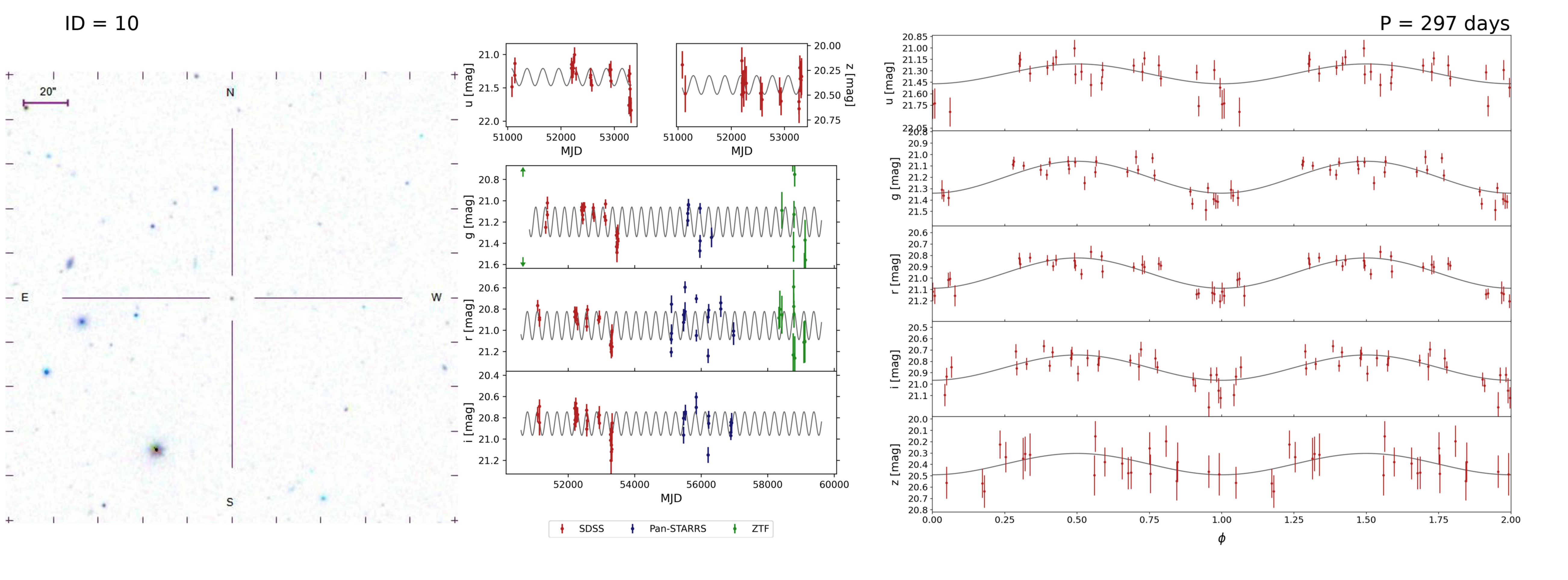}{0.9\textwidth}{a) ID = 10. The variability of this source at later times can not be excluded with PS1 data. ZTF data are too noisy to give definitive arguments. \label{fig:Q_10}}}
	\gridline{\fig{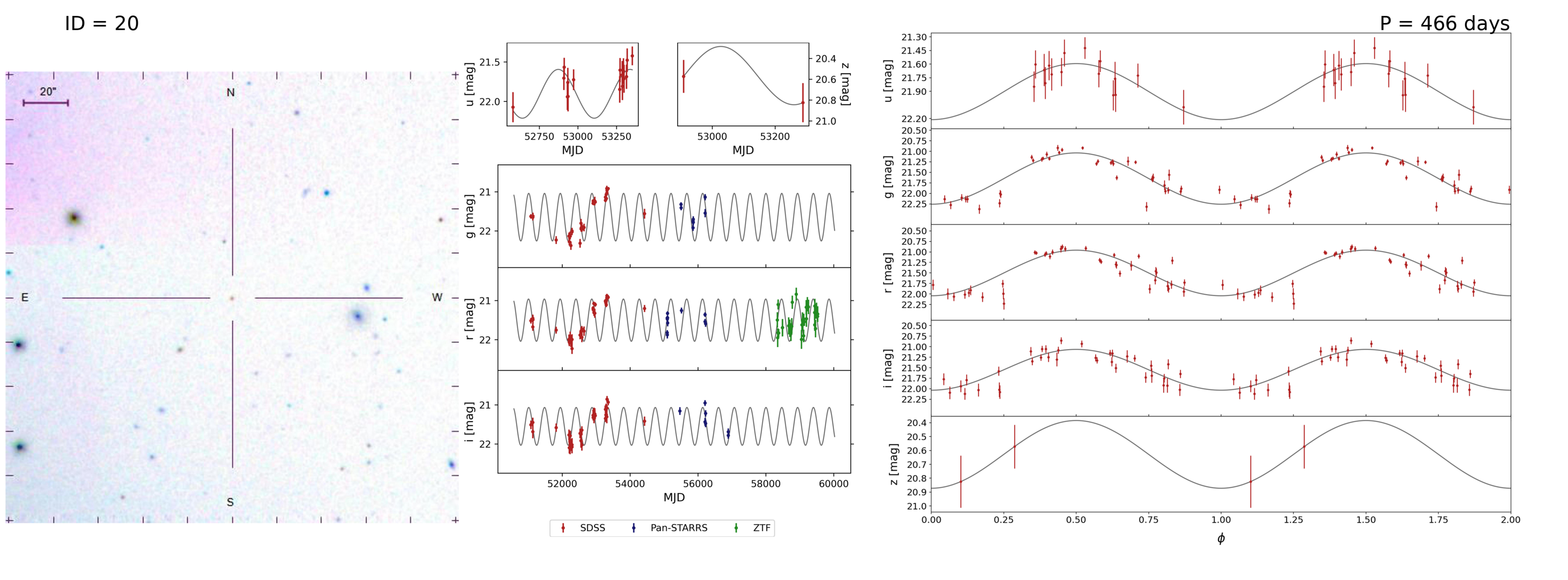}{0.9\textwidth}{b) ID = 20. This source has a confirmation od the variability both from ZTF (only in \textit{r} band) and from PS1 (\textit{gri} bands). Also, the possibility of periodicity cannot be ruled out. \label{fig:Q_20}}}
	\gridline{\fig{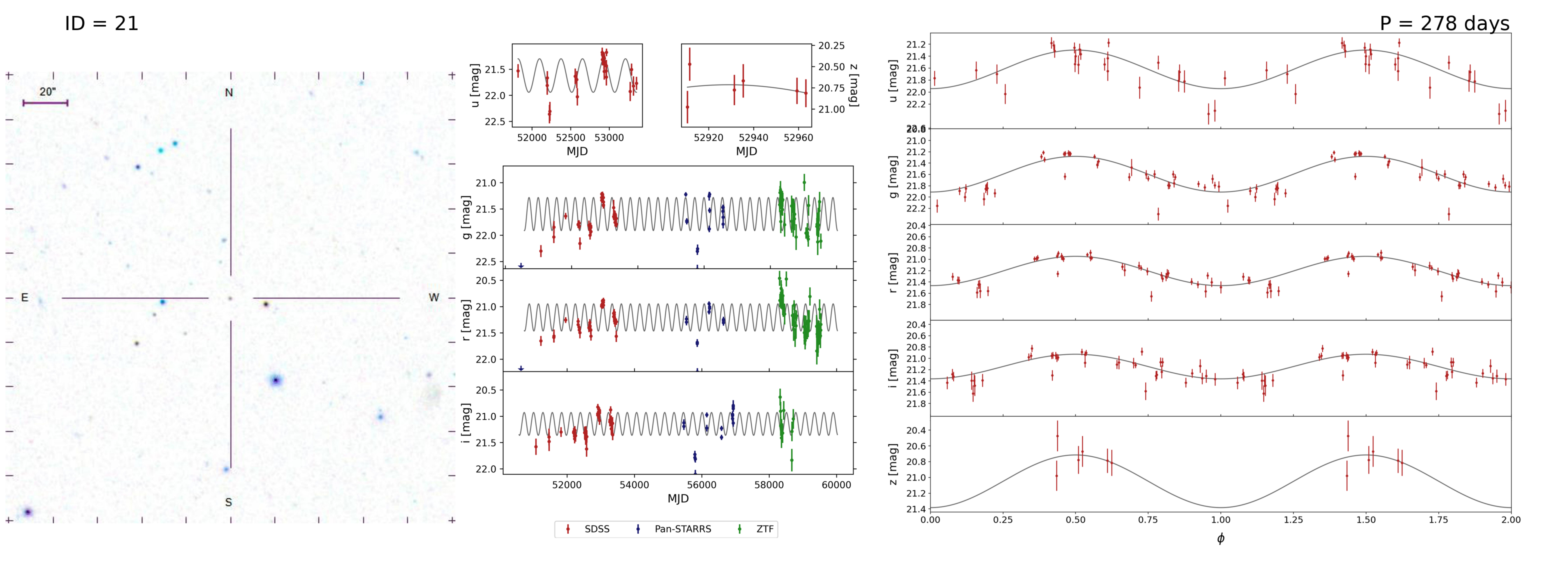}{0.9\textwidth}{c) ID = 21. Although there is a lot of noise in the ZTF data, the observations at later times can not rule out peridical variability of this source. PS1 has much fewer data points, but it still shows variable behavior. \label{fig:Q_21}}}

	\caption{Variable candidates. }
	\label{fig:qsos_1}}
\end{figure*}

\clearpage

\newpage
\begin{figure*}[h!]
	\centering
	\gridline{\fig{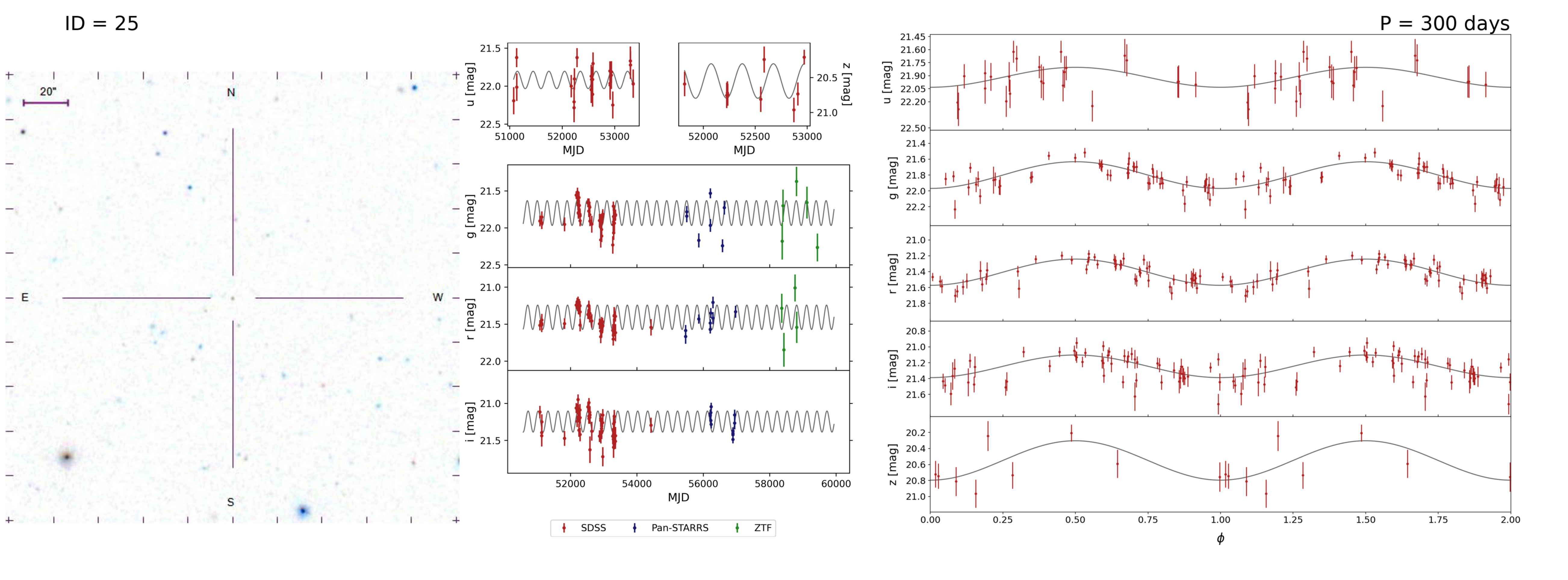}{0.90\textwidth}{d) ID = 25. Due to too few points and too much noise, we cannot conclude anything about the ZTF, but PS1 data show variable behavior. \label{fig:Q_25_1}}}
	\gridline{\fig{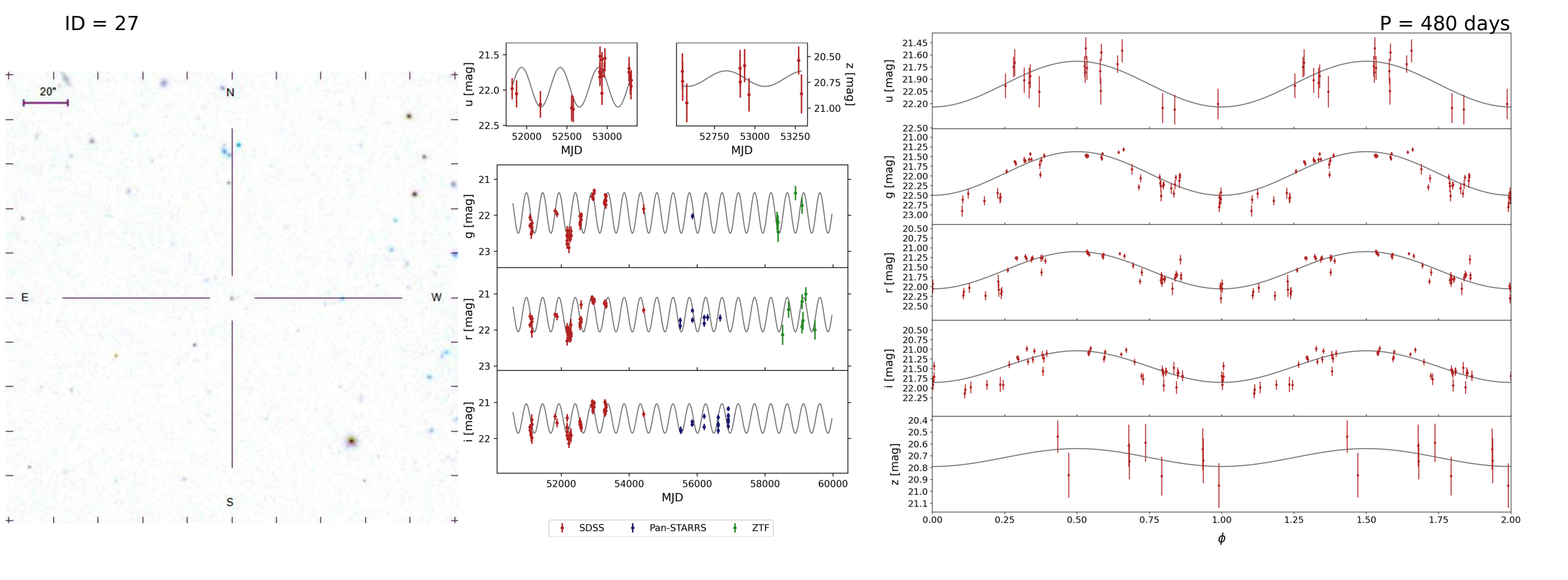}{0.90\textwidth}{f) ID = 27. ZTF data in \textit{gr} bands can not rule out the possibility of periodic behavior for this source at later times. PS1 data are inconclusive or show lower amplitude variability. \label{fig:Q_27}}}
	\caption{Variable candidates.}
	\label{fig:qsos_2}
\end{figure*}

\noindent
\clearpage
\newpage
\pagebreak
\noindent

\section{Results of Monte Carlo simulations} \label{Sec:AppB}

The reliability of the chosen period-finding procedure has been verified by comparing the derived periods to the multiband model and by employing Monte Carlo simulations. To this end, we have chosen to use the multiband extension of the Lomb-Scargle algorithm. 

We have chosen the algorithm implemented in the \textit{gatspy} package \citep{2015ApJ...812...18V}. The multiband periodogram extends the Lomb-Scargle period finding algorithm by treating multiband observations as a dataset with a categorical variable for the observed band. The model uses a single fitting procedure for all observed bands, constructed by including a set of Lomb-Scargle model parameters for each band. 

We have employed Monte Carlo simulations to estimate the reliability of the derived periods.  
We base our simulations on the best-fitting gatspy light curve model of magnitudes $m$, observed at time $t$ and in the band $b$, $m(t,b)$. A particular mock observation, $i$, consists of a simulated point: $(t_i, b_i, m_i, \sigma_{m,i})$. Instead of taking a simple bootstrap that would only reshuffle the dataset, we employed the Gaussian kernel density estimator (KDE) to first determine the  distributions of $t$, $m$, and $\sigma_{m}$, and then to draw a random point from these distributions.  This procedure produced mock lightcurves that consisted of the same number of points as the original dataset. The data points were assumed to be independently distributed among the different bands, with each band appearing as many times in the mock lightcurve as in the original lightcurve. When choosing the optimal KDE distributions, we found that the magnitude error distribution is best reproduced when estimating $\log\sigma_{m}$ using KDE instead of estimating the KDE of $\sigma_{m}$ (see Tisanić et al., in prep). The magnitudes at a particular mock time were computed by fitting the multiband Lomb-Scargle function to the original dataset, computing its prediction at the mock time (and for the chosen randomly-selected band), and then by adding to this value the value of the simulated error $\sigma_{m}$.

We drew 1000 random lightcurves per object and set the number of mock times to the number of observing times in the original light curves. The \textit{gatspy} multiband periodogram has been run on each mock dataset to estimate the reliability of the derived periods.  We employed the robust method of deriving the standard deviations of the period distributions using the 16th and the 84th percentiles of the simulated period distributions. We then computed the P-values based on the estimated sigma value and our periods fitted using the \textit{astropy} implementation of the Lomb-Scargle periodogram derived in the main body of this work, labeled as expected P-values in Table~\ref{Tab:tab2}. Since a portion of the standard deviations were smaller than one day, we capped the standard deviations used for computing the P-values to 0.1 days.

\begin{table*}[h]
             
	\caption{Results of Monte Carlo simulations of the sample's periods. \label{Tab:tab2}  }  
	\centering     
  
	\resizebox{\textwidth}{!}{
		\begin{tabular}{ccccccc}    
			\hline\hline  
			ID&$\sigma_{P}$&expected period with errors&gatspy period with errors&simulated period with errors&expected P-value ($\sigma\to\mathrm{max}(\sigma,0.1d$))&gatspy P-value ($\sigma\to\mathrm{max}(\sigma,0.1d$)) \\    
						\hline
			10&2&297 $\pm$ 2&296 $\pm$ 2&296 $\pm$ 2&0.66&0.99 \\
			20&2&466 $\pm$ 2&467 $\pm$ 2&466 $\pm$ 2&0.73&0.56 \\
			21&1&278 $\pm$ 1&278 $\pm$ 1&278 $\pm$ 1&0.60&0.99 \\
			25&2&300 $\pm$ 2&302 $\pm$ 2&302 $\pm$ 2&0.21&0.92 \\
			27&3&480 $\pm$ 3&479 $\pm$ 3&478 $\pm$ 3&0.45&0.56 \\
			\hline     
		\end{tabular}
	}
	\tablecomments{The results of Monte Carlo simulations based on the \textit{gatspy} multiband algorithm used to infer the reliability of the \textit{astropy}-based period-finding procedure. After the \textit{gatspy}-derived period for each source has been determined, we produced a set of 1000 mock lightcurves, as described in detail in \ref{Sec:AppB}. In short, the mock observations were simulated by estimating the underlying time, magnitude, band and magnitude error distributions from the KDE fit to the real observations. Each mock lightcurve was then fitted using the same \textit{gatspy} algorithm, yielding a set of mock sample periods. The derived distribution of mock sample periods was used to infer the error bars on the expected period - the expected \textit{astropy}-based period calculated, as described in Sect. \ref{sec:initial}. The columns show the following: the internal ID of each source (labeled as `ID'), error bars estimated using Monte Carlo simulations (labeled as '$\sigma_P$'). The periods are listed as follows: the expected period, the \textit{gatspy} best-fitting period, and the mean period derived using Monte Carlo simulations, all with the aforementioned error bars. Additionally, P-values using a capped sigma ($\sigma_{0.1\,\mathrm{d}}=max(\sigma,0.1\,\mathrm{d})$) are listed for the expected- and \textit{gatspy}-derived periods.}
\end{table*}

\newpage
\clearpage
\section{Results of 2D Hybrid method} \label{Sec:AppC}
In general random fluctuations account for a significant portion of the
variance in time series, so that the amplitudes of these stochastic
effects are seen to be greater on longer periods. Due to the fact that the
spectral density of red noise is inversely related to frequency, the red noise
has a particularly significant impact on the lower frequencies. Thus to cope
with this challenge, we also applied time domain periodicity search called 2D
Hybrid method which relies on different types of wavelets
\citep{2018MNRAS.475.2051K, 2019ApJ...871...32K}.
Given two time series $y_{t}$ and $y^{\prime}_{t}$, we can compare their
wavelet matricies (scalograms) $\mathcal{S}$ and $\mathcal{S}^{\prime}$ in
order to know if they follow similar patterns. Our 2D Hybrid method uses
correlation as a  comparison of scalograms \citep{2020OAst...29...51K}.
The 2D Hybrid approach employs various wavelets, e.g. continous, discrete,
Weighted Wavelet Z-transform-WWZ \citep{1996AJ....112.1709F}, high-resolution superlets \citep{moca2021time},  and both observed light curves and
their models. The method generates a contour map of the intensity of (auto)
correlation  on a period-period plane defined by two independent period axes
matching to the two time series (or one). The map is symmetric and can be
integrated along any  of the axes, yielding in a periodogram-like curve of
the strength of correlation among oscillations \citep[for more details
see][]{2018MNRAS.475.2051K, 2019ApJ...871...32K}.

The significance of a detected period $\sigma_P$, we calculated by shuffling of
time series \citep{2019MNRAS.484...19J} so that the period was recomputed
over this new modified data set and the height of the maximum peak in 2D
Hybrid integrate map was compared to that found for the original simulated
data. This process was repeated N (e.g. 100, as the wavelet computation)
times and the significance level was then determined as
\citep{2019MNRAS.484...19J}:
\begin{equation}
\sigma_{P}=\frac{x}{N}
\end{equation}
where $x$ represents the number of times that the peak power of the period
in the original data was greater than that of the uncorrelated ensemble.

We also used another approach for estimating significance which is based
on moving block bootstrap (MBB) methodology
\citep{10.23943/princeton/9780691151687.001.0001, 2012ada..confE..16S,
	2018ApJS..236...16V}. In MBB apporach, blocks of data of a given length are
glued together to create a new time series. Similarly to shuffling method, we
first calculated periods of bootstraped mock light curves.
Then the Generalized extreme value (GEV)  distribution is fitted to the
histogram of detected peaks in mock curves. For a given range of significance
levels (p-values) we obtain the associated confidence level from the fitted
GEV.
If the candidate period has a peak value greater than the confidence level
then we can reject the null hypothesis that the observed light curve is not
periodic with given significance level.

Applying our 2D Hybrid method which core is the WWZ, we obtained period of 
$278.36^{57.34}_{-25.21}$ days, and significance above 99\% measured by
shuffling method, whereas GEV approach produces significance of 90\%.
In order to apply the 2D hybrid method with the high-resolution superlets core, we need
homogenous data. Thus we modeled observed light curve with Deep Gaussian process
\citep[see
e.g.][]{deepGPs13} consisting of latent variable layer and two  gaussian
process layers. The Gaussian processes have  Matern$_{12}$ and Matern$_{32}$
kernels. Detected period is $288.62^{+28.84}_{-07.35}$ days, with significances
of 95\% based on shuffling method and 88\% based on MBB.

\bibliography{ref}{}
\bibliographystyle{aasjournal}

\end{document}